\documentstyle[preprint,aps]{revtex}
\tightenlines
\begin{document}
\draft
\title
{A note on exclusion statistics parameter and Hausdorff dimension}
\author
{Wellington da Cruz\footnote{E-mail: wdacruz@fisica.uel.br}} 
\address
{Departamento de F\'{\i}sica,\\
 Universidade Estadual de Londrina, Caixa Postal 6001,\\
Cep 86051-970 Londrina, PR, Brazil\\}
\date{\today}
\maketitle
\begin{abstract}
We obtain for an anyon gas in the high temperature limit a 
relation between the exclusion statistics parameter $g$ and the
 Hausdorff dimension $h$, given by $g=h(2-h)$. The 
 anyonic excitations are classified into equivalence 
classes labeled by Hausdorff dimension, $h$, and in that limit,
 the parameter $g$ give us the second virial coefficient 
 for any statistics, $\nu$. The anyonic excitations into the same 
 class $h$ get the same value of this virial coefficient. 
 
\end{abstract}



We have obtained in\cite{R1} a distribution function for anyonic 
excitations classified into equivalence classes labeled by 
the fractal parameter $h$, the Hausdorff dimension. We have that 
 for, $h=2$ we get bosons and for 
$h=1$ we get fermions. For, 
$1$$\;$$ < $$\;$$h$$\;$$ <$$\;$$ 2$, we have anyonic 
excitations which interpolates between these two extremes. 
The fractal parameter $h$ is related 
 to statistics $\nu$, in the 
  following way:  

\begin{eqnarray}
\label{e.2}
&&h-1=1-\nu,\;\;\;\; 0 < \nu < 1;\;\;\;\;\;\;\;\;
 h-1=\nu-1,\;\;\;\;
\;\;\;\;\;\;\; 1 <\nu < 2;\;\nonumber\\
&&h-1=3-\nu,\;\;\;\; 2 < \nu < 3;\;\;\;\;\;\;\;\;
h-1=\nu-3,\;\;\;\;\;\;\;\;\;\;\; 3 < \nu < 4;\;\nonumber\\
&&h-1=5-\nu,\;\;\;\; 4 < \nu < 5;\;\;\;\;\;\;\;\;
h-1=\nu-5,\;\;\;\;\;\;\;\;\;\;\; 5 < \nu < 6;\;\\
&&h-1=7-\nu,\;\;\;\; 6 < \nu < 7;\;\;\;\;\;\;\;\;
h-1=\nu-7,\;\;\;\;\;\;\;\;\;\;\; 7 < \nu < 8;\;\nonumber\\
&&h-1=9-\nu,\;\;\;\;8 < \nu < 9;\;\;\;\;\;\;\;\;
h-1=\nu-9,\;\;\;\;\;\;\;\;\;\; 9 < \nu < 10;\nonumber\\
&&etc,\nonumber
\end{eqnarray}

\noindent such that this spectrum of $\nu$ shows us a 
complete mirror symmetry. Now, we obtain 
a connection between the statistics $\nu$ and 
the exclusion statistics 
parameter $g$
\cite{R2} for an anyon gas in the high temperature 
limit, as follows:
  
\begin{eqnarray}
\label{e.3}
&&g=\nu(2-\nu), \;\;\;\;\;\;\;\;\;\;\;\;\;\;\;\; 0 < \nu < 2
;\nonumber\\
&&g=(\nu-2)(4-\nu), \;\;\;\;\;\;\; 2 < \nu < 4;\nonumber\\
&&g=(\nu-4)(6-\nu), \;\;\;\;\;\;\; 4 < \nu < 6;\\
&&g=(\nu-6)(8-\nu), \;\;\;\;\;\;\; 6 < \nu < 8;\nonumber\\
&&g=(\nu-8)(10-\nu), \;\;\;\;\; 8 < \nu < 10;\nonumber\\
&&etc.\nonumber
\end{eqnarray}

\noindent We observe that in \cite{R3} $g$ was obtained only 
for the first interval and this parameter is proportional, 
in general, to the dimensionless second virial coefficient 
if the equation of state of the system admits a virial 
expansion in the high temperature limit. In this note, we 
follow an approach completely distinct and extend that 
result putting it under a new perspective. The expressions 
Eq.(\ref{e.3}) were possible because of the mirror symmetry 
as just have said above. On the other hand, our approach 
in terms of Hausdorff dimension, $h$\cite{R1} collect into 
equivalence class the anyonic 
excitations and so, we consider on equal footing the 
excitations in the class. Therefore, we have 
established independently of the approach given 
 in\cite{R3} the relation between $g$ and the second virial 
 coefficient for the complete spectrum of $\nu$, that is, 
 we have found that 
 the exclusion statistics parameter $g$ is related to $h$, 
 for an anyon gas in the high temperature limit, as follows
 
 \begin{equation}
 \label{e.10}
  g=h(2-h).
 \end{equation}

 \noindent We can check 
 Eq.(\ref{e.10}) using the relations 
 between $h$ and $\nu$ Eq.(\ref{e.2}), obtaining the 
 expressions Eq.(\ref{e.3}). 
 On the other hand, as the second virial coefficient, 
 ${\cal B}_{2}(g)$, determines the 
 exclusion statistics parameter\cite{R3}, we have that
 
\begin{eqnarray}
\label{e.8}
&&{\cal B}_{2}(g)=\frac{g}{2}-\frac{1}{4};\\
&&\tilde{{\cal B}}_{2}(h)=\frac{h(2-h)}{2}-\frac{1}{4},\nonumber
\end{eqnarray}

\noindent where the second expression follows 
from our approach and it is more general, in the sense that, 
the second virial coefficient is written now for each class 
of the anyonic excitations. The expressions Eq.(\ref{e.8}) 
have the correct values for bosons, 
${\cal B}_{2}(0)=-\frac{1}{4}=\tilde{{\cal B}}_{2}(2)$ and for fermions,
${\cal B}_{2}(1)=\frac{1}{4}=\tilde{{\cal B}}_{2}(1)$. In this way, 
we see that the anyonic excitations into the class $h$ have 
the same value for the second virial coeficient in the 
high temperature limit.

\end{document}